\documentclass[conference]{IEEEtran}

\IEEEoverridecommandlockouts
\usepackage{cite}
\usepackage{amsmath,amssymb,amsfonts,mathtools} 
\usepackage{algorithm, algpseudocode}
\usepackage{graphicx}
\usepackage{textcomp}
\usepackage{xcolor}
\usepackage{bm}
\usepackage{float}
\usepackage{soul}
\usepackage[acronym,shortcuts]{glossaries}
\usepackage{outlines}
\usepackage{algorithm}
\usepackage[english]{babel}
\usepackage{blindtext}
\usepackage{subcaption}
\usepackage{orcidlink}
\usepackage{lipsum}
\usepackage{stfloats}
\usepackage[]{hyperref}

\def\BibTeX{{\rm B\kern-.05em{\sc i\kern-.025em b}\kern-.08em
T\kern-.1667em\lower.7ex\hbox{E}\kern-.125emX}}

\newacronym{PDF}{PDF}{probability density function}
\newacronym{CR}{CR}{cognitive radio}
\newacronym{SotA}{SotA}{state of the art}
\newacronym{OFDM}{OFDM}{orthogonal frequency division multiplexing}
\newacronym{PAPR}{PAPR}{peak-to-average power ratio}
\newacronym{IFFT}{IFFT}{inverse fast Fourier transform}
\newacronym{DFT}{DFT}{discrete Fourier transform}
\newacronym{SQNR}{SQNR}{signal-to-quantization-noise ratio}
\newacronym{ADC}{ADC}{analog to digital converter}
\newacronym{DAC}{DAC}{digital to analog converter}
\newacronym{TR}{TR}{tone reservation}
\newacronym{RT}{RT}{reserved tone}
\newacronym{PRT}{PRT}{peak-reserved tone}
\newacronym{LP}{LP}{linear programming}
\newacronym{SLM}{SLM}{selective mapping}
\newacronym{DVB-T2}{DVB-T2}{DVB-second generation}
\newacronym{DVB-NGH}{DVB-NHG}{DVB for next generation Handled}
\newacronym{QCQP}{QCQP}{quadratically constrained quadratic program}
\newacronym{FP}{FP}{fractional programming}
\newacronym{CCDF}{CCDF}{complementary cumulative distribution function}
\newacronym{CDF}{CDF}{cumulative distribution function}
\newacronym{PMF}{PMF}{probability mass function}
\newacronym{IM}{IM}{index modulation}
\newacronym{JCAS}{JCAS}{joint communication and sensing}
\newacronym{ATSC}{ATSC}{advanced television systems committee}
\newacronym{ISAC}{ISAC}{Integrated Sensing and Communication}
\newacronym{FD}{FD}{frequency-domain}
\newacronym{TD}{TD}{time-domain}
\newacronym{QAM}{QAM}{quadrature amplitude modulation}
\newacronym{PSK}{PSK}{phase shift keying}
\newacronym{SNR}{SNR}{signal-to-noise ratio}
\newacronym{wlg}{wlg}{without loss of generality}
\newacronym{ACF}{ACF}{auto-correlation function}
\newacronym{A-ACF}{A-ACF}{aperiodic ACF}
\newacronym{Rx}{Rx}{receiver}
\newacronym{CP}{CP}{cyclic prefix}
\newacronym{RPE}{RPE}{radar parameter estimation}
\newacronym{ISI}{ISI}{inter-symbol interference}
\newacronym{MA}{MA}{multiple access}
\newacronym{PSL}{PSL}{peak to sidelobe level}

\begin{document}
%
\title{Tone Reservation-Based PAPR Reduction Using Manifold Optimization for OFDM-ISAC Systems\\[-0.75ex]}
\author{\IEEEauthorblockN{Getuar Rexhepi\textsuperscript{\orcidlink{0009-0002-3268-522X}},
Kuranage Roche Rayan Ranasinghe\textsuperscript{\orcidlink{0000-0002-6834-8877}}
\\and Giuseppe~Thadeu~Freitas~de~Abreu\textsuperscript{\orcidlink{0000-0002-5018-8174}}}
\IEEEauthorblockA{\emph{School of Computer Science and Engineering},\\ \emph{Constructor University}, Bremen, Germany \\
(grexhepi,kranasinghe,gabreu)@constructor.university\\[-5.75ex]}
\and
\IEEEauthorblockN{David~Gonz{\'a}lez~G.\textsuperscript{\orcidlink{0000-0003-2090-8481}}}
\IEEEauthorblockA{\emph{Wireless Communications Technologies},\\ \emph{Continental AG}\\
Frankfurt/Main, Germany\\
david.gonzalez.g@ieee.org\\[-5.75ex]}
}

\maketitle 

\begin{abstract}
We consider the \ac{PAPR} reduction challenge of \ac{OFDM} systems utilizing \ac{TR} under a sensing-enabling constraint, such that the signals placed in the \acp{RT} can be exploited for \ac{ISAC}. 
To that end, the problem is first cast as an unconstrained manifold optimization problem, and then solved via an iterative projected gradient descent algorithm assisted by an approximation of the infinity norm.
Simulation results show that the proposed method, while maintaining a level of \ac{PAPR} reduction similar to \ac{SotA}, not only has lower computational complexity but also outperforms the alternatives in terms of sensing performance. 
\end{abstract}

\glsresetall
%
\vspace{-3ex}
\section{Introduction}
%
%
\Ac{OFDM} is one of the most commonly used modulation techniques in modern wireless communication systems, due to its robustness against multipath fading, its ability to reduce \ac{ISI} and its seamless integration with \ac{MA} schemes \cite{Molisch2011}. 
However, a well-known drawback of \ac{OFDM} is the high \ac{PAPR} \cite{ParedesG15}, whose significance increases with the number of subcarriers involved in the \ac{IFFT} carried out during the modulation procedure. 
A large \ac{PAPR} is particularly undesirable in \ac{OFDM} systems since it degrades the \ac{SQNR} of both the \ac{ADC} and \ac{DAC} while also significantly reducing the efficiency of the power amplifiers at the transmitter \cite{Udaigiriya2015}.

The goal of \ac{PAPR} reduction techniques is to lower the distance between the signal's peak and average in the time domain without distorting its frequency spectrum.
Over the past two decades, many \ac{PAPR} reduction techniques have been developed to address this issue in \ac{OFDM} systems including coding schemes, clipping, constellation shaping, \ac{SLM} and \ac{TR} \cite{WenIWCMC2008,Tellado1998, JiangTB2008, RahmatallahCST2013, SelahattinTVT2019, TarakTB2021}.

Amongst these methods, the \ac{TR} technique \cite{Tellado1998} -- whereby a small portion of the \ac{OFDM} subcarriers, commonly known as the \acp{PRT}, are reserved not for data transmission but to carry random signals designed to reduce the \ac{PAPR} of the overall transmitted \ac{TD} \ac{OFDM} signal -- stands out for its simplicity.
Originally formulated as a convex optimization problem, the \ac{TR} problem can be efficiently solved by casting it into a \ac{LP} framework \cite{Tellado1998}.

Due to its simplicity and effectiveness, this framework has been adopted by various standards, including \ac{DVB-T2} \cite{DVB-T2Standard2012}, \ac{DVB-NGH} and the \ac{ATSC} 3.0 standard \cite{ATSC3_Standard2024} of the advanced television systems committee.
However, in \ac{SotA} \ac{PAPR} reduction methods based on the \ac{TR} approach \cite{LvMEC2013, LiTWC2020, BulusuITBC2018}, all the \ac{PRT} subcarriers are used  solely for the purpose of \ac{PAPR} reduction. 

Motivated by the emergence of \ac{ISAC} technologies \cite{SturmJPROC2011}, the research community has started to consider the \ac{PAPR} reduction problem in conjunction with the integration of sensing capabilities in the \ac{OFDM} waveform \cite{XiaoyanArxiv2021, RubingWCNC2023}.
Very little work exists, however, on such integration of \ac{ISAC} and \ac{PAPR} reduction in a manner to take advantage of the well-know and effective \ac{TR} approach.
In fact, to the best of our knowledge, the only work in that direction so far is  \cite{VarshneyTRS2023}, where was shown that it is possible to exploit the \ac{PRT} subcarriers for the purpose of sensing.

As demonstrated in literature for sensing scenarios \cite{HeTSP2009}, the unimodularity of the transmit signal is a key performance indicator for the detection/\ac{RPE} abilities of such a system.  
Moreover, to avoid masking weak targets by the echoed signals of strong targets, waveforms with low autocorrelation sidelobes are preferred \cite{Stoica2012}, subsequently requiring efficient algorithms for their exploitation.

In light of the above discussion, the proposed method is based on designing unimodular reserved tones for \ac{PAPR} reduction in \ac{OFDM} systems, thus enabling sensing capabilities via the reserved tones at the mono-/bistatic\footnote{For simplicity and due to the lack of space, we only elaborate on the performance in a monostatic setting equipped with full duplex capabilities to mitigate the inherent self-interference in this manuscript as done in \cite{RanasingheICASSP2024} since the bistatic setting requires some additional considerations such as a connected fronthaul to exploit the indices of the reserved tones.} receiver.
However, as a result of the unimodularity constraint limiting the search space of the reserved tones composed of complex random signals to the unit circle, the optimization problem becomes non-convex and thus, more challenging to solve.

To circumvent this strong limitation, we cast the \ac{PAPR} reduction problem onto a manifold optimization framework to formulate an unconstrained manifold optimization problem on the unit circle in the complex plane and solve it via a projected gradient descent algorithm \cite{Absil2007}.
Simulation results indicate that the proposed method achieves quasi-optimal \ac{PAPR} reduction performance,  with a lower computational complexity compared to the existing optimal methods, while also enabling sensing capabilities based on the reserved tones.
In addition, we compare the sensing and \ac{PAPR} reduction performance of our method with the optimal reserved tones that achieve the best \ac{PAPR} reduction performance utilizing the \ac{QCQP} technique.

The rest of the paper is structured as follows:
\begin{itemize}
\item First, the system model encompassing the \ac{OFDM} signal and radar channel models with the appropriate definitions and metrics is described in Section \ref{sec:system_model}.
\item Next, the classical \ac{SotA} \ac{TR} optimization framework for \ac{PAPR} reduction is presented in Section \ref{sec:PAPR_red_using_TR}.
\item Subsequently, the proposed sensing-enabled \ac{PAPR} reduction method using \ac{TR} and the manifold optimization framework are introduced in Section \ref{sec:sensing_enabled_PAPR}.
\item Finally, a thorough performance analysis of the proposed method including the simulation parameters, the \ac{PAPR} reduction evaluation and the sensing performance evaluation is provided in Section \ref{sec:performance_analysis}. 
\end{itemize}

%
%
\section{System Model}
\label{sec:system_model}
%
%

\subsection{OFDM Signal Model}

We consider a typical \ac{OFDM} system with $N$ subcarriers in the \ac{FD}, such that the discrete \ac{TD} transmit signal $\mathbf{x}$ corresponding to an information vector of complex symbols chosen from an arbitrary constellation ($e.g.$ \ac{QAM}) $\bm{x}\!\in\!\mathbb{C}^{N\times 1}$ can be expressed as
\begin{equation}
\label{eq:OFDM}
\mathbf{x} = \mathbf{F}_N^{\mathrm{H}} \cdot \bm{x} \;\; \in \;\; \mathbb{C}^{N\times 1},
\end{equation}
where $\mathbf{F}_N \!\in\!\mathbb{C}^{N\times N}$ denotes the normalized \ac{DFT} matrix of size $N$ and $(\cdot)^\mathrm{H}$ defines the conjugate transpose operation.

\subsection{Radar Channel Model}

Under the paradigm of monostatic \ac{RPE} for \ac{OFDM} systems and following \cite{Braun2014}, the element-wise input-output relationship for an arbitrarily sampled \ac{TD} signal passing through a radar channel is expressed as
\begin{equation}
\label{eq:RadarChannel}
y[n] = \sigma_X \cdot \sum_{u=1}^U x[n - \tau_u] \cdot e^{j2\pi n \nu_u} + z[n] \in \mathbb{C},
\end{equation}
where $y[n]$, $x[n]$ and $z[n]$ are the $n$-th elements of the received signal, transmitted signal {$\mathbf{x}\!\in\!\mathbb{C}^{N \times 1}$ given in equation \eqref{eq:OFDM}} and the complex Gaussian variable with zero mean and unit variance modelling the noise, respectively, with $\sigma^2_X$ denoting the \ac{SNR}.
Subsequently, the round-trip time delay $\tau_u \in [0,\tau_\text{max}]$ and Doppler shift $\nu_u \in [-\nu_\text{max},+\nu_\text{max}]$ for each $u$-th target is respectively defined as
\begin{equation}
\label{eq:TimeDelay}
\tau_u = \frac{2d_u}{c}, \;\;\;\;\; \text{and} \;\;\;\;\; \nu_u = \frac{2v_u}{\lambda},
\end{equation}
with $d_u \in [0,d_\text{max}]$ denoting the distance between the radar transceiver and the $u$-th target; $v_u \in [-v_\text{max},+v_\text{max}]$ is the velocity of the $u$-th target; $c$ is the speed of light and $\lambda$ is the wavelength of the transmitted signal $\mathbf{x}$.

It is worth noting that due to the carrier frequency and bandwidth of a given system, the maximum time delay $\tau_\text{max}$ and maximum Doppler shift $\nu_\text{max}$ are consequently bounded by the maximum distance $d_\text{max}$ and maximum velocity $v_\text{max}$, respectively.

For simplicity and \ac{wlg}, only static targets are considered in this manuscript, resulting in $\nu_u = 0, \forall u$.
Therefore, the radar parameter estimation problem resolves to estimating the range of the targets, which can then be calculated by estimating the corresponding time delays $\tau_u$.

\subsection{PAPR Definition}

The associated instantaneous \ac{PAPR} measures the maximum power of an arbitrary \ac{TD} signal $\mathbf{x}$ composed of $N$ discrete samples relative to its average, and is defined as \cite{DinurTC2001}
\begin{equation}
\label{eq:PAPR}
\text{PAPR}(\mathbf{x}) \triangleq \frac{\|\mathbf{x}\|_{\infty}^2}{\frac{1}{N}\| \mathbf{x} \|_2^2},
\end{equation}
where $\| \cdot \|_{\infty}$ and $\| \cdot \|_2$ denote the $\ell_{\infty}$ and $\ell_2$ norm operations, respectively.

\subsection{Sensing Performance Metric: Aperiodic ACF (A-ACF)}
The \ac{ACF} of a signal is a crucial performance metric for range estimation, particularly during the matched filtering process at the receiving end. 
The ACF can be characterized as either the linear or periodic self-convolution of the signal, depending on whether a \ac{CP} is included. 
For the purposes of this discussion and \ac{wlg}, we will focus solely on the aperiodic case where no CP is applied \cite{Liu2024}, keeping in mind that the extension to a \ac{CP} inclusive case is straightforward.
The \ac{A-ACF} of a signal $\mathbf{x}$ is defined as
\begin{equation}
r_{k} \triangleq \mathbf{x}^{H} \mathbf{J}_{k} \mathbf{x} = r_{-k}^{*}, \quad k = 0,1, \ldots, N-1,
\end{equation}
where \(\mathbf{J}_{k}\) is the \(k\)-th shift matrix computed via
\begin{equation}
\mathbf{J}_{k} \triangleq \left[\begin{array}{cc}
\mathbf{0} & \mathbf{I}_{N-k} \\
\mathbf{0} & \mathbf{0}
\end{array}\right].
\end{equation}

Exploiting the symmetry of the \ac{ACF} also yields
\begin{equation}
\mathbf{J}_{-k} \triangleq \mathbf{J}_{k}^{T}=\left[\begin{array}{cc}
\mathbf{0} & \mathbf{0} \\
\mathbf{I}_{N-k} & \mathbf{0}
\end{array}\right],
\end{equation}
where $(\cdot)^\mathrm{H}$ denotes the matrix transpose operation.

%
%
\section{PAPR Reduction using Tone Reservation}
\label{sec:PAPR_red_using_TR}
%
%

\subsection{Classical Optimization Framework}

Following the most recent \ac{SotA} work on \ac{PAPR} reduction \cite{LvMEC2013,LiTWC2020,BulusuITBC2018}, the \ac{TR} method relies on the reservation of a certain number of subcarriers, which instead of carrying information symbols, are instead allocated dummy signals designed to minimize the \ac{PAPR} of \ac{OFDM} transmit signals.
Mathematically, this statement corresponds to the fact that the \ac{FD} signal $\bm{x}$ in equation \eqref{eq:OFDM} can be decomposed as
\begin{subequations}
\label{eq:TR_x}
\begin{equation}
\label{eq:TR_x_FD}
\bm{x} = \bm{d} + \bm{r},
\end{equation}
such that the equivalent \ac{TD} representation is
\begin{equation}
\label{eq:TR_x_TD}
\mathbf{x} = \mathbf{d} + \mathbf{r},
\end{equation}
where $\bm{d}\!\in\!\mathbb{C}^{N\times 1}$ and $\mathbf{d}\!\in\!\mathbb{C}^{N\times1}$ correspond to information-carrying communication symbols in the \ac{FD} and \ac{TD}, respectively, while $\bm{r}\!\in\!\mathbb{C}^{N\times 1}$ and $\mathbf{r}\!\in\!\mathbb{C}^{N\times1}$ correspond to the \acp{RT} in the respective domains and the equivalence between equations \eqref{eq:TR_x_FD} and \eqref{eq:TR_x_TD} follows from the linearity of the \ac{IFFT} and the fact that $\mathbf{d} = \mathbf{F}_N^{\mathrm{H}}\bm{d}$ and $\mathbf{r} = \mathbf{F}_N^{\mathrm{H}}\bm{r}$.
\end{subequations}

In order to avoid the overlap of data and \acp{RT}, a mutually-exclusive subcarrier allocation is necessary.
To that end, let $\mathcal{N}\!\triangleq\!\{1,\dots,N\}$ denote the set of all subcarrier indices, such that the sets of data and reserved tone subcarrier indices can be respectively denoted by $\mathcal{D}\subseteq \mathcal{N}$ and $\mathcal{R}\subseteq \mathcal{N}$, with $\mathcal{N} = \mathcal{D}\cup\mathcal{R}$ and $\mathcal{D}\cap\mathcal{R} = \emptyset$.
Let us also define the corresponding cardinalities $N_d \triangleq |\mathcal{D}|$ and $N_r \triangleq |\mathcal{R}|$, with $N \!=\! N_\mathrm{D} + N_\mathrm{R}$.

Denoting the $d$-th element of $\bm{d}$ and the $r$-th element of $\bm{r}$ positioned at the $n$-th location in $\bm{x}$ to be $d_n$ and $r_n$, respectively, it follows that the data and \ac{RT} vectors are such that their entries satisfy
\begin{equation}
d_n  =  0, \forall\, n \in \mathcal{R},\quad\text{and}\quad
r_n  = 0, \forall\, n \in \mathcal{D}.
\end{equation}

\subsection{SotA QCQP Technique}

For later convenience, we introduce the notation $\bm{r}\in\mathbb{C}^\mathcal{R}$ to indicate that the elements of the sparse vector $\bm{r}$ corresponding to the indices in $\mathcal{R}$ take on complex numbers, while the remaining entries are zero, such that the optimization problem for \ac{PAPR} minimization can then be concisely formulated via a \ac{QCQP} framework via
\begin{equation}\label{refeq:QCQP}
\begin{aligned}
& \underset{\bm{r} \in \mathbb{C}^{N \times 1}}{\text{minimize}}
& & \| \mathbf{d} + \mathbf{F}_N^{\mathrm{H}} \bm{r} \|_\infty^2, \\
& \text{subject to}
& & \| \bm{r} \|^2  \leq {P_{\max}}, \\
& & & r_i = 0, i \in \mathcal{R}^c, \\
\end{aligned}
\end{equation}
where $P_\mathrm{\max}$ is a power constraint, $\mathcal{R}$ is the set of reserved tones that are orthogonal to the data carrying tones, which in turn is the complement of the set of data carrying tones, $\mathcal{D}$, i.e. $\mathcal{R} = \mathcal{D}^{c}$.
To reduce the number of computations in the optimization procedure, let us also define a modified \ac{IFFT} matrix $\mathbf{F}_R^{\mathrm{H}} \in \mathbb{C}^{N \times N_r}$ with each column corresponding to the position of a \ac{RT} in the \ac{OFDM} signal. 
Consequently, the \ac{FD} \ac{RT} vector now can be defined as $\bm{r} \in \mathbb{C}^{N_r \times 1}$.

\section{Sensing-enabled PAPR Reduction using TR}
\label{sec:sensing_enabled_PAPR}

\subsection{Problem Formulation}
Leveraging the reformulated notation in Section \ref{sec:PAPR_red_using_TR}, the new optimization problem can be formulated as
\begin{equation}
\label{eq:first_form}
\begin{aligned}
& \underset{\mathbf{r} \in \mathbb{C}^{N_r \times 1}}{\text{minimize}}
& & \| \mathbf{d} + \mathbf{F}_R^{\mathrm{H}} \bm{r} \|_\infty^2, \\
& \text{subject to}
& & \| \bm{r} \|^2 \leq {P_{\max}}, \\
& & & |r_i| = 1, i = 0 \dots N_r-1, \\
\end{aligned}
\end{equation}
where the constraint $|r_i| = 1$ ensures that the reserved tones are located on the unit complex circle enabling theoretically perfect detection properties \cite{HeTSP2009}.
Consequently, since the power is always equal to  $\| \bm{r} \|^2 = N_r$ due to the unit circle constraint, equation \eqref{eq:first_form} becomes

\begin{equation}
\begin{aligned}
& \underset{\bm{r} \in \mathbb{C}^{N_r \times 1}}{\text{minimize}}
& & \| \mathbf{d} + \mathbf{F}_R^{\mathrm{H}} \bm{r} \|_\infty^2, \\
& \text{subject to}
& & |r_i| = 1, i = 0 \dots N_r-1. \\
\end{aligned}
\end{equation}

\subsection{Reformulated Manifold Optimization Framework}

The optimization problem can now be formulated as an unconstrained manifold optimization problem via

\begin{equation}
\begin{aligned}
& \underset{\bm{r} \in \mathcal{M} }{\text{minimize}}
& & \| \mathbf{d} + \mathbf{F}_R^{\mathrm{H}} \bm{r} \|_\infty^2, \\
\end{aligned}
\end{equation}
where $\mathcal{M}$ is the unit circle on the complex plane.
To solve this manifold optimization problem, we utilize the Riemannian optimization framework, where the optimization problem is executed on the chosen manifold.

\subsection{Proposed Solution: Projected Gradient Descent Algorithm}
Since the unit complex sphere is a smooth manifold, the optimization problem above can be solved using the projected gradient descent algorithm \cite{boumal2023}.
However, the fact that the infinity norm is not differentiable necessitates an approximation via the norm definition as
\begin{equation}
\label{eq:inftynormapprox}
\| \mathbf{x} \|_\infty \triangleq \lim_{p \to \infty} \|\mathbf{x} \|_p,
\end{equation}
where we notice that the squared form of the norm becomes redundant as it does not affect the optimization problem i.e., minimizing $\| x \|_p$ also minimizes $\| x \|_p^2$, where the $p$ norm is defined as
\begin{equation}\label{eq:relaxation}
\| \mathbf{x} \|_p = \left( \sum_{i=1}^{N_r} |x_i|^p \right)^{1/p}.
\end{equation}


Consequently, it is essential to compute the gradient of the $p$-norm for the implementation of our algorithm. Using the approximation in equation \ref{eq:relaxation}, the gradient of the $p$-norm can be computed as
\begin{equation}\label{eq:gradientx}
\frac{\partial}{\partial}_i \| \mathbf{x} \|_p = \left( \frac{|x_i|^{p-2} x_i}{\| \mathbf{x} \|_p^{p-1}}  \right),
\end{equation}
where $\mathbf{x} \triangleq \mathbf{d} + \mathbf{F}_R^{\mathrm{H}} \bm{r}$, while $p$ can be chosen arbitrarily.

As a result, the partial derivative becomes
\begin{equation}\label{eq:grad}
\frac{\partial}{\partial}_i \| \mathbf{d} + \mathbf{F}_R^{\mathrm{H}} \bm{r} \|_p =  \left( \mathbf{F}_R  \frac{|\mathbf{d}_i + \mathbf{F}_R^{\mathrm{H}} \mathbf{r}_i|^{p-2} (\mathbf{d}_i + \mathbf{F}_R^{\mathrm{H}} \bm{r}_i)}{\| \mathbf{d} + \mathbf{F}_R^{\mathrm{H}} \bm{r} \|_p^{p-1}}  \right).
\end{equation}

The final step is a projection of the gradient onto the unit complex circle, which can be achieved via normalizing the \acp{RT} by
\begin{equation}\label{eq:proj}
\bm{r}_{\text{proj}} = \frac{\bm{r}}{|\bm{r} |}.
\end{equation}

A complete summary of the projected gradient descent algorithm is presented in Algorithm \ref{alg:proj_grad}.
The projected gradient descent algorithm is both conceptually straightforward and easy to implement.
Its structure is simple, consisting primarily of iterative gradient updates and projections, which are well-understood and widely applicable.

\addtolength{\topmargin}{0.05 in}
\begin{algorithm}[H]
\caption{Projected Gradient Descent Algorithm}
\label{alg:proj_grad}
\begin{algorithmic}[1]
\State Initialize $\bm{r}$, $\alpha$, $p$, K.
\For{k = 1 \dots K}
\State Compute $\nabla f(\mathbf{c})$ using eq. \eqref{eq:grad}.
\State Update $\bm{r} = \bm{r} - \alpha \nabla f(\bm{r})$.
\State Project $\bm{r}$ onto the unit complex circle using eq. \eqref{eq:proj}.
\EndFor
\end{algorithmic}
\end{algorithm}
\vspace{-2ex}

The core operations of the algorithm, namely, computing the gradient and performing the projection, are computationally efficient and can be implemented in a variety of programming languages.
The algorithm is also highly flexible, as it allows for the use of different norms and step sizes, which can be adjusted to suit the specific requirements of the problem at hand.
In addition, since the objective is a $p$-norm affine function, the objective of the optimization problem is continuous on $\mathbb{C}^n$ satisfying the Lipschitz condition
\begin{equation}
\|  f({\bm{r}}) -  f({\bm{r}}') \| \leq L \| {\bm{r}} - {\bm{r}}' \|, \quad \forall {\bm{r}}, {\bm{r}}' \in \mathbb{C}^n,
\end{equation}
where $L$ is the Lipschitz constant.

Moreover, from\cite{Duistermaat2004}, since the largest step size with guaranteed convergence is given by $\alpha = 1/L$, where $L$ is the Lipschitz constant of the gradient of the objective function, an upper bound can also be given as
\begin{equation}
L \geq \| \nabla f(\bm{r}) \|.
\end{equation} 
Therefore, the objective function is Lipschitz continuous and fast convergence of the algorithm is guaranteed.

\section{Performance Analysis}
\label{sec:performance_analysis}

\subsection{Simulation Parameters}
The algorithm outlined above was tested on a 512 subcarrier \ac{OFDM} system employing 16-PSK with 64 reserved tones.
The parameters chosen can be seen in Table \ref{table:Parameters} and the initial reserved tones $\bm{r}$ can be chosen randomly from the unit circle or can be preset to zero.
The method was implemented in MatlabR2024 and run on a 1.3 GHz Intel Core i5 processor with 16 GB of RAM.
The timing performance can be compared with the optimal reserved tones that achieve the best \ac{PAPR} reduction performance by solving the optimization problem in equation \eqref{refeq:QCQP}.
\vspace{-2ex}
\begin{table}[H]
\centering
\caption{Parameters used for the simulation results.}
\begin{tabular}{|c|c|}
\hline
\textbf{Parameter} & \textbf{Value} \\ \hline
$N$ & 512 \\ \hline 
$N_r$ & 64 \\ \hline
$\alpha$ & 1 \\ \hline
$p$ & 50 \\ \hline
$K$ & 2000 \\ \hline
$R$ & $[1, 9, \dots, 504, 512]$ \\ \hline
\end{tabular}
\vspace{2ex}
\label{table:Parameters}
\centering
\caption{Comparison of different runtimes.}    
\begin{tabular}{|c|c|c|}
\hline
\textbf{Method} & \textbf{Time (s)} & \textbf{PAPR (dB)}  \\ \hline
Optimal (QCQP) & 3.779831 & 4.6896 \\ \hline
Prop. p = 10 & 0.085961 & 5.660204 \\ \hline
Prop. p = 50 & 0.206386 & 5.129761 \\ \hline
Prop. p = 100 & 0.285779 & 5.054547 \\ \hline
Prop. p = 150 & 0.336831 & 5.032135 \\ \hline
\end{tabular}
\label{table:Performance}
\end{table}

\begin{figure}[H]
\centering
\includegraphics[width=\columnwidth]{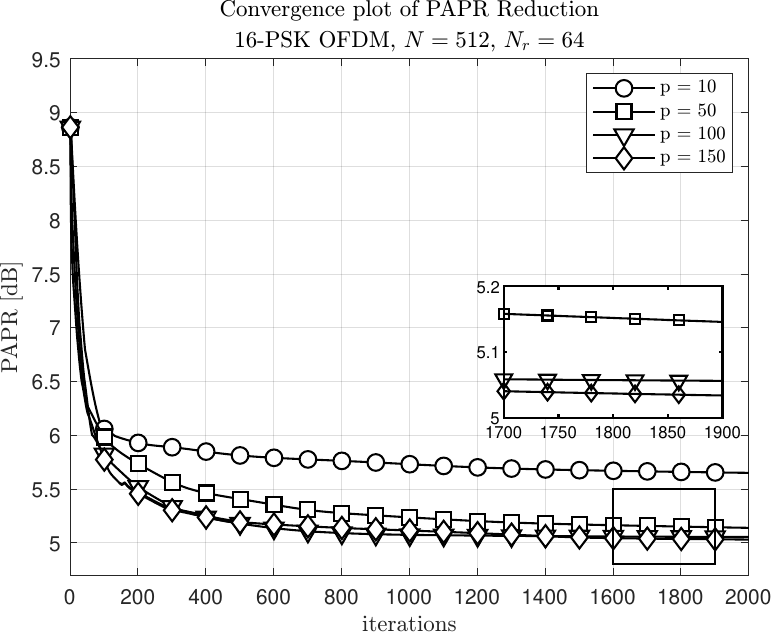}
\vspace{-3ex}
\caption{Convergence of the proposed method for different values of the $p$-norm parameter used in equation \eqref{eq:inftynormapprox}.}
\label{fig:Conv}
\vspace{-2ex}
\end{figure}

The performance of the proposed method is illustrated in Table \ref{table:Performance}, whose numbers should be compared against the \ac{PAPR} of 8.8647 dB corresponding to the \ac{OFDM} system without \ac{PAPR} reduction.
Table \ref{table:Parameters} shows that the proposed method, despite the much lower complexity, achieves a \ac{PAPR} reduction comparable to the optimal \ac{TR} method.
In addition, the quick convergence of the proposed method under the parameters found in Table \ref{table:Parameters} can be seen in Figure \ref{fig:Conv} for different $p$ values.

\subsection{PAPR Reduction Evaluation}

\begin{figure}[H]
\centering
\includegraphics[width=\columnwidth]{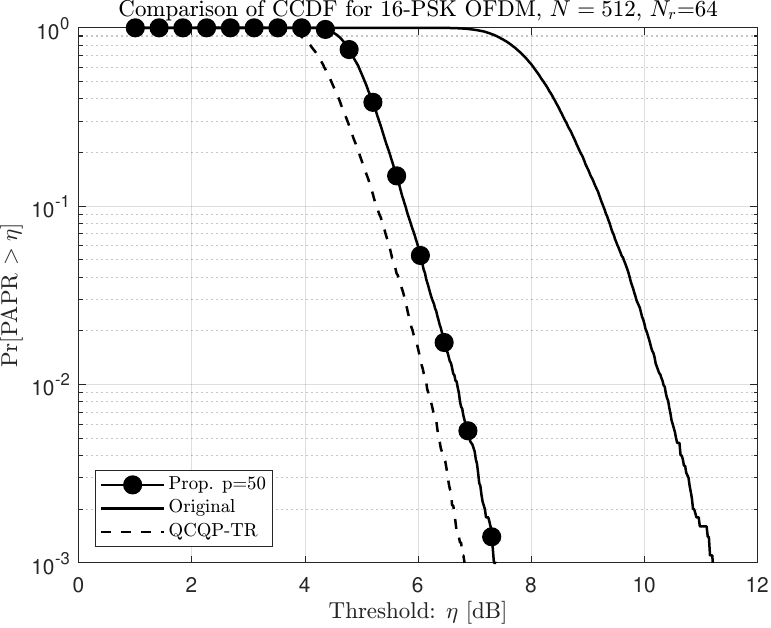}
\vspace{-3ex}
\caption{PAPR reduction performance of proposed method.}
\label{fig:PAPR}
\end{figure}

The \ac{CCDF}, defined as the probability that the \ac{PAPR} of a signal exceeds a certain threshold, is widely used to evaluate the performance of \ac{PAPR} reduction techniques \cite{ChoMIMOOFDM2010}.
The \ac{CCDF} of the proposed method is compared in Figure \ref{fig:PAPR} against those of the optimal \ac{TR}-based method and a system without \ac{PAPR} reduction.

The results show that, under the same constraints, the suggested approach yields a \ac{PAPR} reduction performance that is comparable to the optimal \ac{TR} method.

\subsection{Sensing Performance Evaluation}
To evaluate the detection performance, we can use the radar channel model in equation \eqref{eq:RadarChannel} and the time delay in equation \eqref{eq:TimeDelay}.
For a single target, an \ac{OFDM} system using subcarrier spacing $\Delta f=450$ kHz and carrier frequency $f_c=26$ GHz was considered.
By correlating the received signal with the transmitted signal, we can estimate the time delay, and therefore the range of the target.
From equation \eqref{eq:TimeDelay} we can estimate the distance of the target to be
\begin{equation}
\hat{d}_u = \frac{c \hat{\tau}_u}{2}.
\end{equation}

\begin{figure}[H]
\centering
\includegraphics[width=\columnwidth]{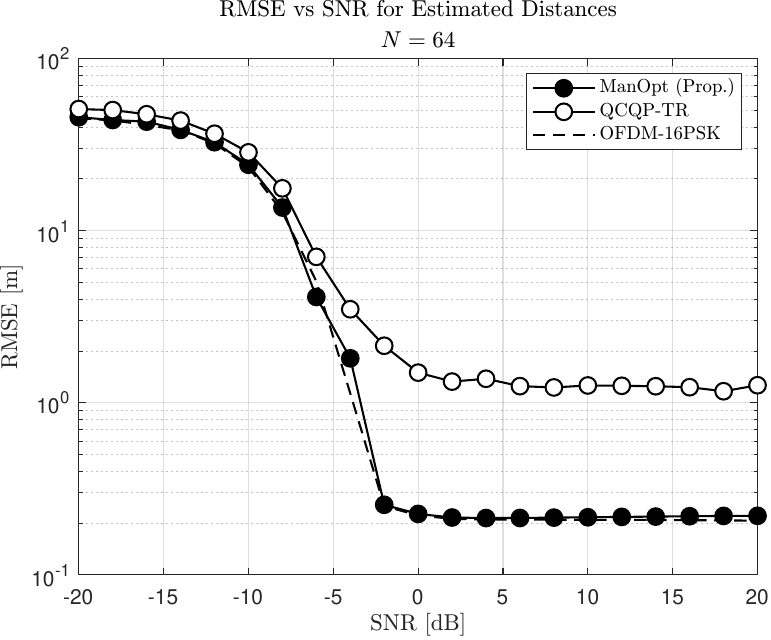}
\caption{One Target Ranging RMSE of different waveforms}
\label{fig:ROC}
\vspace{0.5ex}
\includegraphics[width=\columnwidth]{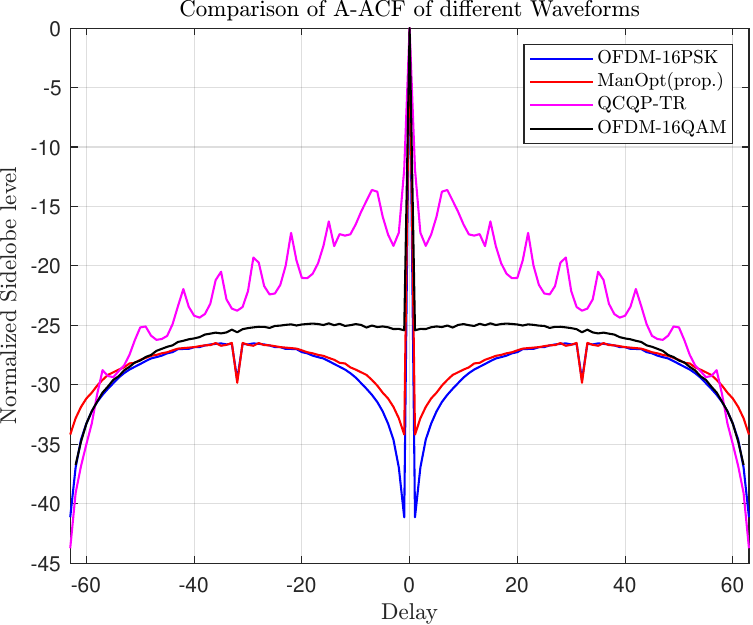}
\caption{Comparison of the A-ACF of different waveforms}
\label{fig:AACF}
\end{figure}

Finally, the radar detection performance of the proposed method can be evaluated as shown in Figure \ref{fig:ROC}.

As observed, utilizing the optimal reserved tones for radar detection results in the worst performance, whereas the proposed method achieves equivalent performance to the 16-\ac{PSK} \ac{OFDM} modulated waveform.
It is noteworthy that Liu et al. \cite{Liu2024} demonstrated that this \ac{OFDM} waveform achieves the lowest sidelobe level in ranging, thereby delivering the best ranging performance.

To further assess the sensing performance, the \ac{A-ACF} of the various waveforms was computed and is presented in Figure \ref{fig:AACF}. 
The \ac{A-ACF} of the proposed method closely matches that of the 16-\ac{PSK} \ac{OFDM}-modulated waveform, which demonstrates the best performance, whereas the optimal reserved tones once again exhibits the worst performance.
In addition, the \ac{PSL} of the proposed method is visibly lower than that of the optimal reserved tones or the 16-\ac{QAM} \ac{OFDM}-modulated waveform, while being comparable to the 16-\ac{PSK} \ac{OFDM}-modulated waveform.
These findings, as anticipated, are therefore consistent with the radar detection performance results.

\section{Conclusions}
%
We address the problem of \ac{PAPR} reduction in \ac{OFDM} systems using \ac{TR} under the assumption of perfect sensing.
The proposed method focuses on the design of unimodular reserved tones for \ac{PAPR} reduction, which simultaneously enable sensing at the monostatic receiver via the reserved tones.
Since the unimodularity constraint limits the search space of the reserved tones composed of complex random signals to the unit circle, the optimization problem becomes non-convex and thus more challenging to solve.
To circumvent this strong limitation, we cast the \ac{PAPR} reduction problem onto a manifold optimization framework to formulate an unconstrained manifold optimization problem on the unit circle in the complex plane and solve it via a projected gradient descent algorithm.
Simulation results indicate that the proposed method achieves quasi-optimal \ac{PAPR} reduction performance, with a lower computational complexity compared to the existing optimal methods, while also enabling sensing capabilities based on the reserved tones.
In addition, we compare the sensing and \ac{PAPR} reduction performance of our method with the optimal reserved tones that achieve the best \ac{PAPR} reduction performance utilizing the \ac{QCQP} technique, which once again highlights the superiority of the proposed technique.
As a direction for future work, we propose extending the method to a bistatic radar scenario and optimizing the indices for allocating the reserved tones to further enhance performance. 

\section{Acknowledgement}

The authors would like to acknowledge the support of Dr. Osvaldo Gonsa as the host  of a research internship at Continental AG during the execution of this work.

\newpage

\end{document}